\documentclass[runningheads]{llncs}

\usepackage[T1]{fontenc}
\usepackage{amsmath}
\usepackage{amssymb}
\usepackage[utf8]{inputenc}
\DeclareUnicodeCharacter{00A0}{~} % Fixes hidden non-breaking spaces

\usepackage{graphicx}

\begin{document}

\title{FMRP-LEAN: A HIPAA-Compliant AI-Augmented LIMS Architecture for End-to-End Clinical Assay Workflow Optimization}

% \titlerunning{Abbreviated paper title}

\author{
Eva McCord\inst{1}\orcidID{0000-0002-9365-4569} \and
Ernest Pedapati\inst{1}\orcidID{0000-0002-7954-5104} \and
Zag ElSayed\inst{2}\orcidID{0000-0001-9094-1469}
}

\authorrunning{E. McCord et al.}

\institute{
Division of Child and Adolescent Psychiatry, Cincinnati Children’s Hospital Medical Center, Ohio, USA \and
School of Information Technology (SoIT), University of Cincinnati, Ohio, USA \\
\email{Eva.McCord@cchmc.org}\\
\email{Ernest.Pedapati@cchmc.org}\\
\email{elsayezs@ucmail.uc.edu}
}

\maketitle

\begin{abstract}
Clinical biomarker workflows in translational research settings often rely on spreadsheet-driven tracking, manual quality control (QC) reconciliation, and loosely integrated systems, resulting in limited state visibility, delayed reporting, and increased operational risk. These challenges are particularly pronounced in multi-day assays such as Luminex-based quantification of Fragile X Messenger Ribonucleoprotein (FMRP), where HIPAA-compliant data governance, deterministic workflow progression, and coordinated communication across laboratory and clinical teams are required.
This paper presents FMRP-LEAN, a HIPAA-compliant, AI-augmented Laboratory Information Management System (LIMS) architecture that formalizes biospecimen lifecycle management through a finite-state workflow model with explicit transition guards and dwell-time observability. The system integrates a self-hosted Supabase/PostgreSQL stack deployed within hospital-controlled infrastructure, hybrid edge–internal isolation with encrypted tunneling and loopback-only services, and bi-directional REDCap synchronization. A unified MRN–UUIDv7 identifier framework with QR-based tracking ensures traceable clinical–research linkage under PHI residency constraints.
FMRP-LEAN incorporates automated statistical QC pre-screening and a governance-constrained AI operations module that operates exclusively on aggregate projections, with deterministic fallback guarantees. Deployment demonstrates improved workflow observability, reduced QC latency, and enhanced cross-role transparency between laboratory technicians, research coordinators, and patient-facing teams. The architecture provides a reproducible model for secure, state-explicit, and AI-augmented clinical research workflows in regulated healthcare environments (\url{https://github.com/drpedapati/fmrplean}).

\end{abstract}

\section{Introduction}
Clinical biomarker quantification workflows in translational research environments frequently rely on fragmented laboratory processes, manual chain-of-custody documentation, and loosely integrated data management systems~\cite{batayah2024smart,bath2011limsportal}. While such workflows may function operationally, they often introduce inefficiencies, delayed turnaround times, increased reviewer burden, and an elevated risk of human error, particularly in multi-day assays that require strict quality control and regulatory compliance~\cite{mitra2016fundamentals}. These limitations are amplified in hybrid clinical research settings where patient-facing communication, research data integrity, and protected health information (PHI) governance must coexist within hospital security boundaries under the Health Insurance Portability and Accountability Act (HIPAA)~\cite{edemekong2018health}.

FMRP dysregulation is central to Fragile X syndrome pathophysiology, and Luminex-based quantification has emerged as a translational biomarker platform requiring multi-day assay coordination and structured reporting~\cite{boggs2022optimization,casingal2020identification,hilal2025dysregulation}. However, the associated assay workflow remains complex and time-intensive, requiring approximately three days per plate and involving dried blood spot (DBS) extraction, incubation cycles, plate preparation, signal detection, statistical quality control (QC), and structured reporting via REDCap~\cite{randol2024variation,le2025research}.

In practice, legacy workflows rely heavily on non-standardized sample identifiers, spreadsheet-based reconciliation, manual QC verification, and delayed cross-team communication. Conventional Laboratory Information Management Systems (LIMS) are often designed either for industrial laboratories or generalized clinical deployments and rarely address the combined requirements of (1) research-specific workflow customization, (2) strict HIPAA-compliant data residency constraints~\cite{hipaa1996}, and (3) safe integration of artificial intelligence (AI) assistance within regulated hospital environments~\cite{yuen2025laboratory}. Furthermore, while AI systems are increasingly introduced into healthcare operations, concerns regarding PHI exposure, auditability, and reliability remain central barriers to deployment~\cite{perez2024mitigating}.

\subsection{Governance-Constrained AI in Regulated Systems}
The integration of large language models (LLMs) into healthcare introduces governance, safety, and regulatory challenges, including privacy leakage, hallucination risk, bias amplification, and limited auditability~\cite{perez2024mitigating,bender2021dangers,weidinger2021ethical}. Regulatory frameworks such as HIPAA, along with guidance from the World Health Organization and NIST, emphasize data minimization, secure system boundaries, transparency, and reliability in AI-assisted clinical environments~\cite{edemekong2018health,ai2023artificial,world2024ethics}. Most healthcare AI systems focus on predictive modeling or decision support using patient-level data~\cite{verma2021implementing}, whereas operational AI augmentation within regulated laboratory workflows remains underexplored.
FMRP-LEAN addresses this gap by enforcing aggregate-only AI prompt construction, ensuring that no MRNs, UUIDs, or record-level data are transmitted outside the protected boundary. AI outputs are strictly advisory and are bounded by a deterministic fallback, in which structured statistical summaries replace narrative generation if the AI service is unavailable. This design reframes AI as an operational co-pilot rather than a clinical decision authority, embedding it within a loopback-restricted, hospital-resident infrastructure under explicit PHI isolation guarantees.
To operationalize this model, FMRP-LEAN integrates a self-hosted Supabase/PostgreSQL stack within the hospital's internal infrastructure, a hybrid edge–internal topology using encrypted tunneling and reverse isolation, REDCap synchronization~\cite{rasyaad2025developing,le2025research}, a unified MRN–UUIDv7 identifier framework with QR-based tracking~\cite{kakolaki2025comparative}, and automated QC pre-screening. Together, these components provide secure, state-explicit workflow orchestration with governance-constrained AI augmentation.

\subsection{Formal PHI Isolation and Deterministic Degradation Model}
Let the clinical research dataset be defined as:
\begin{equation}
D = \{r_1, r_2, \dots, r_n\}.
\end{equation}
Each record $r_i$ consists of:
\begin{equation}
r_i = (\mathrm{MRN}_i, \mathrm{UUID}_i, X_i),
\end{equation}
where $\mathrm{MRN}_i$ represents the clinical identifier (PHI), $\mathrm{UUID}_i$ represents the workflow identifier, and $X_i$ represents assay measurements and associated metadata. 

Define a projection operator $\mathcal{A}(\cdot)$ such that:
\begin{equation}
\mathcal{A}(D) = \{ f_k(D) \mid f_k : D \rightarrow \mathbb{R} \},
\end{equation}
where each $f_k$ computes an aggregate statistic (e.g., counts of pending assays, mean plate coefficient of variation, reagent utilization totals). By construction:
\begin{equation}
\mathcal{A}(D) \cap \{\mathrm{MRN}_i\} = \varnothing,
\end{equation}
and
\begin{equation}
\mathcal{A}(D) \cap \{\mathrm{UUID}_i\} = \varnothing,
\end{equation}
ensuring that no patient-level identifiers or record-specific data are included in the AI prompt input.

The AI prompt construction function is defined as:
\begin{equation}
P = g(\mathcal{A}(D)),
\end{equation}
where $g(\cdot)$ maps aggregate statistics into structured natural-language summaries.

The system enforces the following \textbf{PHI Isolation Constraint}:
\begin{equation}
\forall p \in P, \quad p \not\supseteq \mathrm{PHI},
\end{equation}
meaning that no element of the prompt contains protected health information as defined under HIPAA~\cite{hipaa1996}.

\begin{figure}[htbp]
\centering
\includegraphics[width=0.9\linewidth]{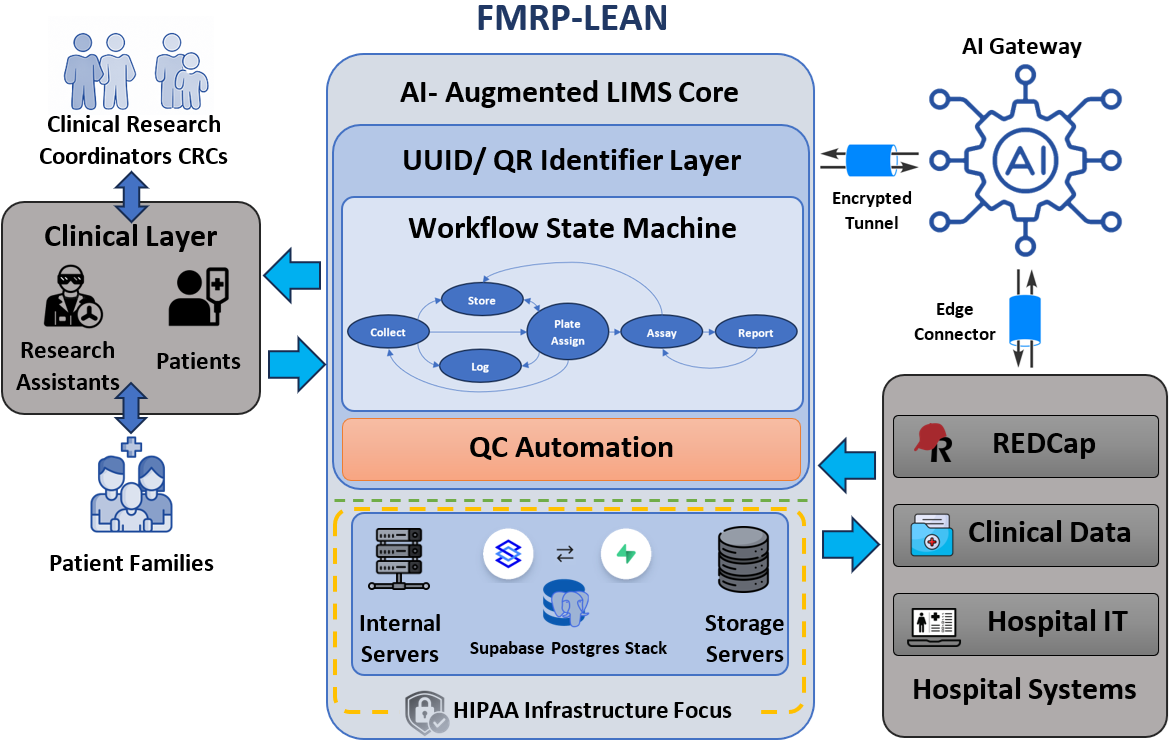}
\caption{FMRP-LEAN HIPAA-compliant LIMS architecture.}
\label{fig1}
\end{figure}

Additionally, the AI output function $h(P)$ is bounded by a deterministic fallback operator:
\begin{equation}
O =
\begin{cases}
h(P), & \text{if AI service available},\\
\mathcal{A}(D), & \text{otherwise}.
\end{cases}
\end{equation}
Thus, system correctness and operational continuity do not depend on AI availability. The AI component serves solely as an augmentative narrative layer over structured aggregates, rather than as a decision-making authority. The conceptual architecture of FMRP-LEAN is shown in Fig.~\ref{fig1}.

The primary contributions of this work are:
\begin{enumerate}
\item A HIPAA-Compliant Hybrid LIMS Architecture.
We present a deployable hybrid edge–internal design that enforces internal PHI residency, loopback-only service binding, encrypted ingress via reverse tunneling, and structured secret management, reducing public attack surface while preserving hospital-controlled accessibility.
\item A Finite-State Workflow Model with Deterministic Observability.
FMRP-LEAN formalizes biospecimen lifecycle progression as a guarded finite-state machine with dwell-time monitoring, replacing spreadsheet-driven tracking with auditable, queryable workflow transitions and measurable backlog visibility.
\item A Governance-Constrained AI Augmentation Framework.
We introduce an aggregate-only projection model for AI input construction with formal PHI isolation and deterministic fallback guarantees, ensuring AI operates as a non-authoritative operational layer without affecting workflow correctness or regulatory compliance.
\item QC Automation with Preserved Dual Verification.
A statistical pre-screening pipeline reduces reviewer burden while maintaining mandatory dual human attestation before reportable state transitions.
\item Cross-Role Transparency in Regulated Clinical Workflows.
Shared state-aware dashboards improve transparency and coordination across laboratory technicians, clinical research coordinators, and patient-facing teams, reducing ambiguity in assay progression and enhancing communication with patient families under strict HIPAA constraints.
\item Real-World Clinical Deployment.
We demonstrate operational improvements in observability, QC latency, auditability, and coordination relative to a legacy spreadsheet-based workflow.
\end{enumerate}

FMRP-LEAN demonstrates that AI augmentation, when architected with explicit security boundaries and workflow formalization, can enhance efficiency, observability, and accountability in regulated clinical research environments. The proposed framework is generalizable to other translational biomarker workflows requiring secure, auditable, and AI-assisted laboratory operations.

\section{Background and Clinical Workflow}
Fragile X Messenger Ribonucleoprotein 1 (FMRP) is a critical regulator of synaptic plasticity, neuronal maturation, and chromatin remodeling~\cite{hilal2025dysregulation,casingal2020identification}. Loss or dysregulation of FMRP due to CGG repeat expansion in the \textit{FMR1} gene is the primary molecular mechanism underlying Fragile X syndrome (FXS), a leading inherited cause of intellectual disability and autism spectrum disorder~\cite{mitra2016fundamentals}. Quantitative measurement of FMRP has emerged as a translational biomarker linking molecular dysfunction to behavioral and neurophysiological phenotypes, including EEG-derived biomarkers and cognitive performance measures~\cite{boggs2022optimization}.

Recent advances in Luminex-based immunoassay techniques have enabled robust detection of FMRP in dried blood spot (DBS) samples with improved reproducibility and clinical utility~\cite{boggs2022optimization}. However, while assay sensitivity has improved, operational complexity remains substantial in real-world clinical research environments.

\subsection{Legacy Assay Workflow and Operational Constraints}
The legacy workflow includes DBS extraction, plate preparation, signal detection, statistical QC, dual-review verification, REDCap upload, and report generation. This pipeline involves multiple handoffs between Clinical Research Coordinators (CRCs) and Research Assistants (RAs), which can create sources of latency and inconsistency. Three primary operational challenges were identified:
\begin{enumerate}
    \item Identifier fragmentation, where legacy workflows relied on redundant or inconsistent label systems, increasing transcription and reconciliation risk.
    \item Inventory opacity, where DBS storage lacked centralized digital tracking, limiting visibility into each sample’s workflow state.
    \item QC reviewer burden, where exported assay data required extensive manual review, contributing to fatigue and delayed turnaround time.
\end{enumerate}
These inefficiencies are not merely administrative, they directly impact turn-around time for patient communication, assay throughput, operational observability, and regulatory traceability.

\subsection{Limitations of Conventional LIMS in Regulated Research Environments}
Commercial Laboratory Information Management Systems (LIMS) are typically optimized for industrial testing laboratories or generalized clinical workflows~\cite{yuen2025laboratory}. While such systems may provide database-backed sample tracking, they often lack (i) flexible workflow state modeling tailored to assay-specific research pipelines, (ii) native or operationally practical REDCap synchronization, (iii) fine-grained HIPAA boundary enforcement in hybrid deployments~\cite{hipaa1996}, and (iv) governance-constrained AI augmentation mechanisms suitable for regulated environments~\cite{perez2024mitigating}.

Moreover, many cloud-based LIMS platforms abd solutions rely on externally hosted databases, which may conflict with institutional requirements for PHI residency within the institution. As a result, a gap exists between research-grade assay workflows, clinical regulatory requirements, and deployable AI-assisted operational optimization. FMRP-LEAN was designed to address this gap by formalizing workflow state transitions, unifying identifiers, centralizing inventory observability, and embedding governance-constrained AI assistance within a strictly controlled hospital infrastructure.

\section{System Architecture}
FMRP-LEAN was designed under five primary architectural constraints: (i) HIPAA-compliant data residency such that participant-level data remain within hospital-controlled infrastructure, (ii) loopback-only service exposure such that core services do not bind to public interfaces, (iii) workflow formalization such that sample lifecycle transitions are explicit and queryable, (iv) PHI-isolated AI augmentation such that AI subsystems operate only over aggregate projections, and (v) deterministic operational continuity, such that system correctness does not depend on AI availability.

\subsection{Layered System Model}
\label{sec:layered_model}
FMRP-LEAN is organized into four logical layers to separate clinical interaction concerns from workflow orchestration, AI augmentation, and infrastructure enforcement.

\subsubsection{Clinical Interaction Layer}
This layer includes CRCs, RAs, and patient/family communication touchpoints. Users interact through authenticated interfaces that support scheduling updates, sample status visibility, and handoff coordination. The interaction layer is intentionally thin to reduce coupling between user-facing workflows and the internal data plane.

\subsubsection{Workflow Orchestration Layer}
\begin{figure}[htbp]
\centering
\includegraphics[width=0.9\linewidth]{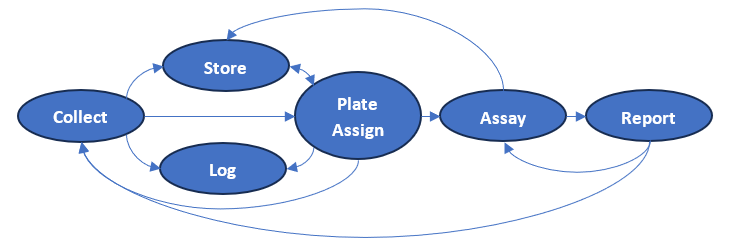}
\caption{Finite-state workflow model for FMRP-LEAN biospecimen lifecycle.}
\label{fig2}
\end{figure}

The workflow layer implements: (i) identifier management and labeling, (ii) explicit workflow state transitions, (iii) centralized DBS inventory tracking, and (iv) pre-QC automation hooks. Each sample is represented by a workflow identifier (UUIDv7), optionally linked to an MRN within the HIPAA boundary. Dashboards compute aggregate state counts and dwell-time summaries to support backlog awareness without exposing PHI. The workflow state machine model is shown in Fig.\ref{fig2}, where Samples transition deterministically across discrete states from collection to reporting. Quality control (QC) validation gates progression, and failed QC results trigger controlled reassignment for reprocessing. This explicit state formalization enables backlog observability and auditability within the HIPAA boundary.

\subsubsection{AI-Augmented LIMS Core}
The LIMS core is implemented as a self-hosted Supabase stack with PostgreSQL as the primary data store and integrated components for authentication, REST access, storage, real-time services, and edge functions~\cite{lorenz2024building}. Public API access is mediated through an API gateway (Kong), while compute extensions (QC ingestion, REDCap sync jobs, and briefing generation) are implemented as internal edge functions.

AI augmentation is implemented as an optional operations module that produces daily laboratory briefings from PHI-safe aggregates. The formal PHI isolation and deterministic degradation model is defined in the PHI model. This design aligns with guidance emphasizing reliability, governance, and safe failure behavior in regulated AI deployments~\cite{ai2023artificial,world2024ethics}.

\subsubsection{Secure Infrastructure Layer}
FMRP-LEAN is deployed using a hybrid edge-internal topology. The internal host operates within the hospital-controlled environment and contains the full Supabase stack and PHI-bearing data store. A public edge host terminates inbound connections and forwards traffic through an encrypted tunnel to internal services bound to loopback interfaces, preventing direct public exposure of database and core service ports while maintaining a stable endpoint~\cite{sisavath2025study}.

\begin{figure}[htbp]
\centering
\includegraphics[width=0.85\linewidth]{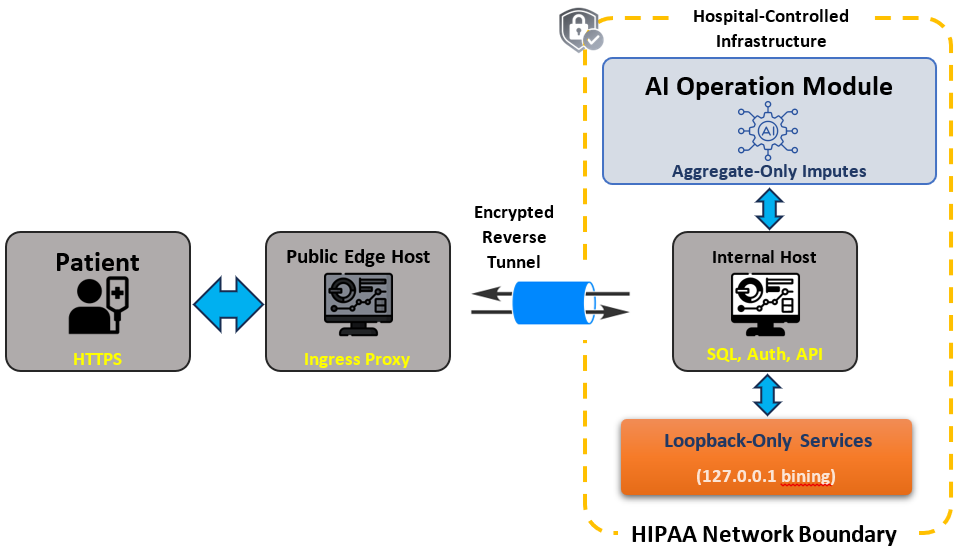}
\caption{Hybrid edge–internal security topology of FMRP-LEAN and PHI isolation structure paradigm.}
\label{fig4}
\end{figure}

The Hybrid edge–internal security topology of FMRP-LEAN, where Client requests terminate at a public edge host and are forwarded through an encrypted tunnel to the internal hospital network. All core services bind exclusively to loopback interfaces within the HIPAA boundary. The AI operations module operates within the protected environment under aggregate-only constraints, shown in Fig.~\ref{fig4}.

\subsection{REDCap Integration Model}
\label{sec:redcap_integration}
REDCap serves as the institutional system of record for participant and visit metadata and supports controlled capture of study-specific research data~\cite{garcia2021research}. FMRP-LEAN integrates with REDCap to (i) ingest participant context required for scheduling and chain-of-custody continuity, and (ii) publish assay status updates and finalized outputs to enable traceable reporting.

Integration is executed via an internal-only runner service deployed inside the protected container network. REDCap credentials are stored outside version control and mounted at runtime, reducing the risk of credential leakage while enabling reliable bi-directional synchronization.

In addition to bi-directional data synchronization, FMRP-LEAN supports participant registration through manual entry or REDCap lookup queries within the protected boundary. A visit calendar module enables structured scheduling of assay visits, while laboratory-facing dashboards expose visit queues and sample processing queues in real time. These features formalize coordination between clinical research staff and laboratory personnel while maintaining PHI residency constraints.

\subsection{Security Boundary Enforcement}
FMRP-LEAN enforces security and compliance through architectural constraints rather than relying solely on administrative controls. PHI-bearing records reside exclusively within internally managed PostgreSQL infrastructure deployed inside the hospital network boundary. Core services (database, auth, REST, storage, and realtime) bind exclusively to loopback interfaces on the internal host, reducing attack surface and preventing unintended public exposure.

Public ingress is terminated at the edge and forwarded through encrypted tunneling, ensuring cross-boundary communication occurs only over authenticated, encrypted channels. Secrets (database credentials, API keys, gateway tokens) are stored outside source control in protected environment files with restricted file permissions. Finally, the AI subsystem is structurally isolated from PHI-bearing records by enforcing aggregate-only input construction.
Edge protection includes HTTPS enforcement and DDoS mitigation through Cloudflare-based ingress shielding. Deployment updates follow rolling release practices with version-controlled rollback capability, allowing rapid reversion in the event of faulty updates.

\subsection{Architectural Novelty and Differentiation}
FMRP-LEAN extends beyond conventional LIMS deployments by combining (i) formal workflow modeling, (ii) hybrid boundary enforcement with loopback-only internal services, and (iii) governance-constrained AI augmentation within a unified architecture. Traditional cloud-hosted LIMS platforms may provide centralized tracking but often rely on external databases and do not enforce explicit PHI isolation for AI modules. On-premise systems preserve residency but often lack encrypted edge-to-internal isolation and bounded AI integration. In contrast, FMRP-LEAN constrains AI to aggregate-only inputs with deterministic fallback and prevents AI outputs from mutating workflow state or assay values. This reframes AI as an operational co-pilot rather than a decision authority, reducing governance risk while delivering operational value.

\section{AI-Augmented, Workflow Optimization and Quality Control Automation}
In the legacy workflow, sample status was implicitly tracked through spreadsheets and ad hoc coordination between CRCs and RAs, limiting visibility into bottlenecks and making backlog prioritization subjective. FMRP-LEAN replaces this with explicit state tracking and real-time observability. Each biospecimen is assigned a UUIDv7 workflow identifier and is tracked across discrete lifecycle states. The system generates PHI-safe operational summaries (state counts, dwell-time distributions, plate capacity usage) that support dashboards and daily operational planning. Beyond operational efficiency, explicit workflow state modeling significantly improves cross-role transparency. Clinical research coordinators, laboratory personnel, and patient-facing teams share a unified, state-aware view of assay progression. Rather than relying on informal status inquiries or spreadsheet reconciliation, stakeholders can observe real-time lifecycle positioning (e.g., Stored, PlateAssigned, QCPending, Reported). This reduces communication latency, mitigates ambiguity, and improves clarity when communicating expected turnaround timelines to patient families.

\subsection{Automated Quality Control Layer}
The Luminex-based assay produces structured exports containing replicate measurements, calibration curve parameters, and derived concentration values. In legacy practice, QC relied on manual inspection and spreadsheet validation. FMRP-LEAN introduces semi-automated QC pre-screening executed after assay export. Statistical predicates include coefficient-of-variation thresholds, dynamic range checks, and replicate outlier detection. Samples failing checks are flagged for enhanced manual review. Importantly, QC automation does not replace human oversight. Transition into verified/reportable states remains gated by dual independent human attestation. Automation serves as a pre-filter, reducing reviewer burden while preserving clinical-grade rigor.

\subsection{Governance-Constrained AI Operations Module}
The AI operations module generates daily briefings intended to improve coordination and situational awareness. AI inputs are restricted to aggregate operational metrics (e.g., state counts, mean dwell time per state, QC anomaly counts, plate utilization, reagent utilization). No MRNs, UUIDs, or record-level attributes are included, satisfying the PHI isolation constraint. AI output is advisory and does not modify workflow state transitions, assay measurements, or QC decisions. If AI is unavailable, deterministic fallback returns the structured aggregates directly, ensuring operational continuity and auditability.

Beyond daily operational briefings, the AI module includes a Plate Planning Assistant that generates PHI-safe optimization suggestions for assay plate allocation based on aggregate workload statistics and capacity constraints. These recommendations remain advisory and do not directly modify workflow state.

\subsection{System-Level Optimization Effects}
By formalizing workflow state transitions, enforcing dwell-time prioritization, and embedding bounded AI augmentation, FMRP-LEAN transforms a manually coordinated assay pipeline into a measurable, auditable system without relaxing compliance constraints. The dashboard also supports AI-assisted research productivity summaries derived from aggregate processing metrics.

\section{Security, Compliance, and Risk Analysis}
Operational resilience is supported through encrypted ingress protection, DDoS mitigation at the edge, HTTPS transport security, tunnel auto-restart policies, health-check monitoring, rolling deployment updates, and version-controlled rollback mechanisms~\cite{perez2024mitigating,bender2021dangers,world2024ethics}. Logical exports of workflow queues and storage records provide additional administrative backup safeguards for regulatory archiving.

\section{Evaluation and Deployment Analysis}
FMRP-LEAN was deployed within a hospital-controlled environment supporting Luminex-based FMRP quantification. The evaluation compares system behavior with the pre-deployment legacy workflow, which relied on spreadsheet-based tracking, manual QC reconciliation, informal prioritization, and ad hoc status communication between laboratory staff and clinical research coordinators (CRCs).

\subsection{Operational Outcomes}
\begin{figure}[htbp]
\centering
\includegraphics[width=0.86\linewidth]{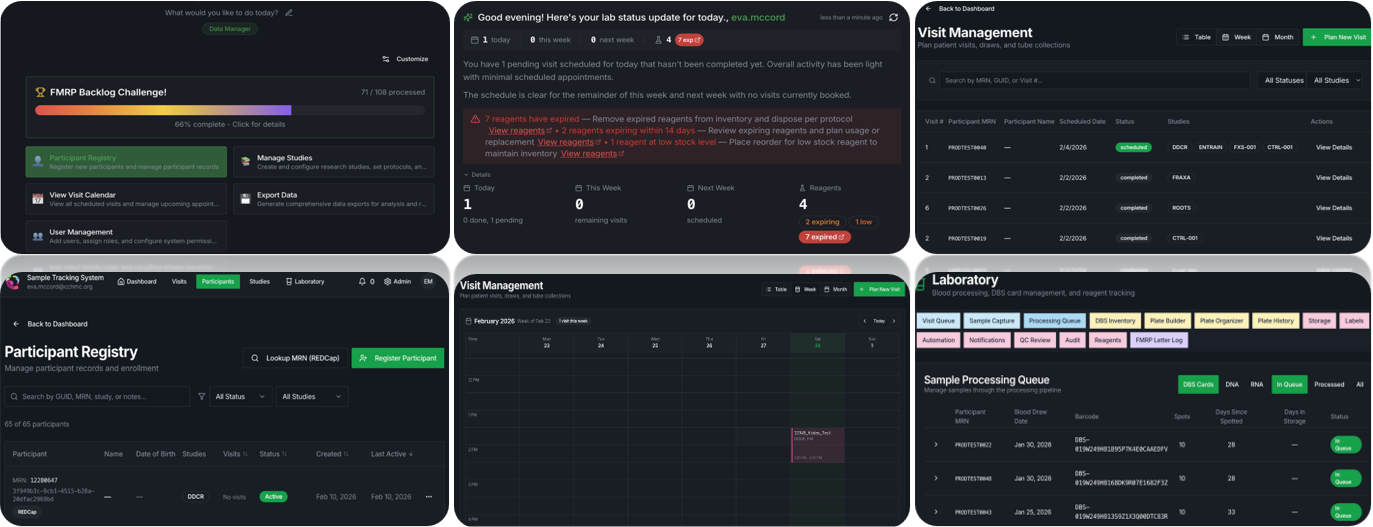}
\caption{Sample screenshots fo the FMRP-LEAN Implementation system operation Mode.}
\label{fig6}
\end{figure}

Explicit finite-state modeling replaced spreadsheet tracking, enabling synchronized state counts and dwell time summaries. Automated statistical pre-screening reduced reviewer burden while preserving dual attestation. Deterministic dwell time prioritization reduced variability in sample progression and improved turnaround consistency. Representative snapshots of the system interface, illustrating state-aware dashboards and QC pre-screening outputs, are shown in Fig.~\ref{fig6}.

\subsection{Evaluation Baseline}
\begin{figure}[htbp]
\centering
\includegraphics[width=0.75\linewidth]{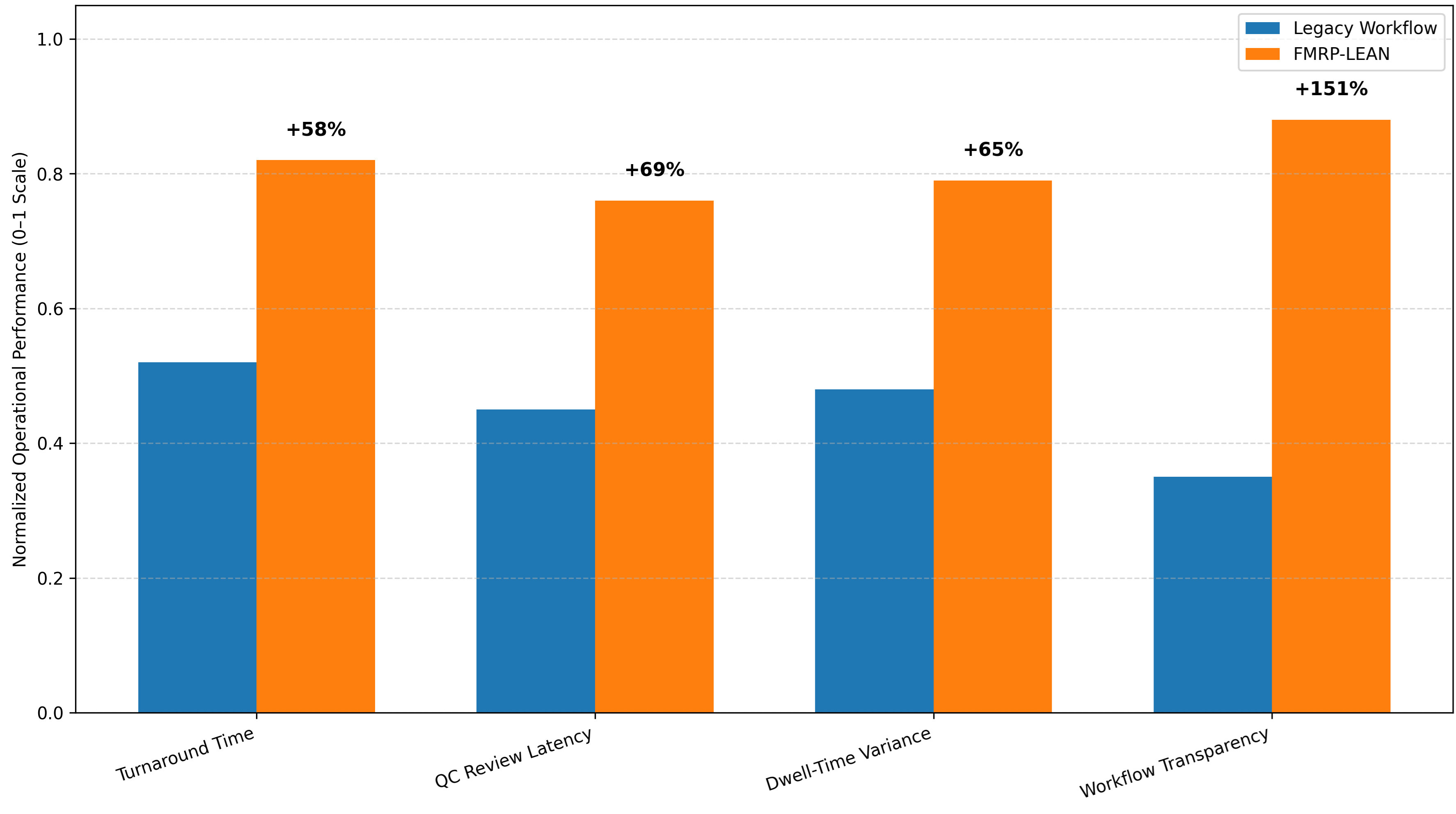}
\caption{Before \& after comparison of clinical assay workflow performance following deployment of FMRP-LEAN.}
\label{fig5}
\end{figure}

System performance was evaluated across four operational dimensions:
\begin{enumerate}
    \item Workflow Observability: real-time visibility into sample lifecycle states.
    \item Quality Control Latency: time from assay export to verified state.
    \item Backlog Management Consistency: variance in dwell time within backlog-prone states.
    \item Cross-Role Transparency: clarity and coordination across laboratory technicians, CRCs, and patient-facing teams.
\end{enumerate}
Where institutional constraints limit disclosure of raw metrics, normalized improvements are illustrated in Fig.~\ref{fig5}.

\subsection{Cross-Role Transparency}
Deployment improved cross-role transparency by replacing manual status reconciliation with shared state-aware workflow visibility. Laboratory personnel and CRCs observe synchronized lifecycle progression, enabling clearer communication with patient families regarding assay status and expected timelines without compromising PHI residency.

\subsection{Architectural Implications}
Operational improvements resulted from workflow formalization, PHI containment, deterministic prioritization, and bounded AI augmentation rather than changes to biological assay procedures. These results demonstrate that transparency and efficiency gains in regulated clinical laboratories can be achieved through architectural rigor and governance-aware design.

\subsection{Comparison to state-of-the-art}
To further contextualize these results relative to prevailing LIMS deployment models, Table \ref{table1} compares FMRP-LEAN with representative cloud-hosted and traditional on-premise systems across governance, workflow formalization, and AI integration dimensions.

\begin{table}[h]
\caption{Architectural and governance comparison of FMRP-LEAN relative to representative cloud-hosted and on-premise LIMS models.}
\label{table1}
\centering
\begin{tabular}{|p{4.9cm}|p{2.5cm}|p{2cm}|p{1.35cm}|}
\hline
\textbf{Feature} & \textbf{Cloud LIMS} & \textbf{On-Prem LIMS} & \textbf{FMRP-LEAN} \\
\hline
Internal PHI Residency & Often external & Yes & \textbf{Yes} \\
\hline
Loopback-only service exposure & Rare & Rare & \textbf{Yes} \\
\hline
Formal workflow state model & Limited & Limited & \textbf{Yes}\\
\hline
Operational REDCap integration & Limited & Limited & \textbf{Yes} \\
\hline
AI augmentation & Rare & No & \textbf{Yes} \\
\hline
Aggregate-only AI inputs & No & No & \textbf{Yes} \\
\hline
Deterministic AI fallback & No & No & \textbf{Yes} \\
\hline
\end{tabular}
\end{table}

While conventional LIMS platforms provide centralized sample tracking, they typically do not enforce explicit PHI isolation constraints for AI modules, loopback restricted service binding, or deterministic AI fallback guarantees. FMRP-LEAN integrates these safeguards within a unified architecture, enabling measurable workflow transparency and operational improvements without relaxing regulatory constraints.

\section{Discussion and Limitations}
\subsection{Generalization Beyond FMRP Assays}
Although developed for Luminex-based FMRP quantification, the core abstractions, finite-state modeling, dwell-time prioritization, QC pre-screening, and aggregate only AI summaries are assay-agnostic.
Most healthcare AI deployments target diagnostic prediction or decision support using patient-level inputs~\cite{rajkomar2019machine}. In contrast, FMRP-LEAN constrains AI to PHI-safe aggregate projections and prohibits AI outputs from mutating workflow state or assay values. This bounded operational model offers a lower-risk pathway for AI adoption in regulated environments, aligning with governance frameworks emphasizing transparency, reliability, and safe failure modes~\cite{ai2023artificial,world2024ethics}. The architecture reframes AI as an operational co-pilot rather than a clinical decision authority.

\subsection{Limitations}
This work represents a single-site observational deployment. Broader validation across institutions and assay types is needed to confirm generalizability. While PHI isolation is enforced architecturally, sustained compliance relies on institutional governance practices, including credential rotation and periodic audit.

\section{Conclusion}
This paper presented FMRP-LEAN, a HIPAA-compliant, AI-augmented LIMS architecture designed to optimize end-to-end clinical assay workflows within a regulated hospital environment. FMRP-LEAN integrates workflow formalization, hybrid edge-internal boundary enforcement, encrypted ingress isolation, and governance-constrained AI augmentation in a unified system.
A key contribution is the formal PHI isolation and deterministic degradation model, which bounds AI inputs to aggregate-only projections and guarantees workflow continuity under AI unavailability. Deployment demonstrates improved workflow observability, enhanced assay traceability, reduced QC overhead, and improved coordination between clinical and research personnel.
Beyond Fragile X biomarker analysis, the architectural abstractions introduced in FMRP-LEAN are generalizable to other translational workflows requiring secure, auditable, and governance-aware AI augmentation in healthcare.
The framework demonstrates that regulated clinical biomarker workflows can be made transparent and state-explicit across laboratory and patient-facing roles under strict PHI constraints.

\bibliographystyle{splncs04}
\bibliography{references}

\end{document}